\documentclass[12pt]{article}

\input epsf
\usepackage{amssymb,wrapfig}
\usepackage[matrix,arrow,curve]{xy}
\usepackage{cite}
\usepackage{graphicx}

\def\aa {{\cal A}}
\def\nn {{\cal N}}
\def\rr {{\mathbb R}}
\def\cc {{\mathbb C}}
\def\hh {{\mathbb H}}
\def\pp {{\mathbb P}}
\def\zz {{\mathbb Z}}

\def\ka {K\"ahler}

\newcommand{\nc}{\newcommand}
\nc{\eps}{\epsilon}

\begin{document}

    \begin{titlepage}

    \begin{center}

    \vskip .3in \noindent

    {\Large \bf{A simple class of $\nn=3$ gauge/gravity duals}}

    \bigskip

    Daniel Louis Jafferis$^1$ and Alessandro Tomasiello$^2$\\

    \bigskip

    $^1$ Department of Physics, Rutgers University, Piscataway, NJ 08855, USA\\
    $^2$ Jefferson Physical Laboratory, Harvard University,
    Cambridge, MA 02138, USA

    \vskip .5in
    {\bf Abstract }
    \vskip .1in
    \end{center}

    \noindent

We find the gravity duals to an infinite series of $\nn=3$ Chern--Simons quiver theories.
They are AdS$_4\times M_7$ vacua of M--theory,
with $M_7$ in a certain class of 3--Sasaki--Einstein manifolds obtained by a quotient
construction. The field theories can be engineered from a brane configuration;
their geometry is summarized by a ``hyperK\"ahler toric fan'' that can be read off
easily from the relative angles of the branes. The singularity at the tip of the cone
over $M_7$ is generically not an orbifold. The simplest new manifolds we consider
can be written as the biquotient ${\rm U}(1)\backslash {\rm U}(3)/{\rm U}(1)$.
We also comment on the relation between our theories and
four--dimensional $\nn=1$ theories with the same quiver.

    \vfill
    \eject


    \end{titlepage}

\section{Introduction} 
\label{sec:intro}

The AdS/CFT correspondence is at present best understood for
AdS$_5$. Shortly after the original proposed duality for
AdS$_5\times S^5$ \cite{maldacena}, which is dual to an $\nn=4$
CFT$_4$, some simple $\nn=1$ generalizations
were put forward \cite{kachru-silverstein,klebanov-witten}.
The correspondence was later applied to more general Calabi--Yau cones;
today nice combinatorial methods exist
\cite{hanany-kennaway,franco-hanany-kennaway-vegh-wecht,
hanany-vegh,feng-he-kennaway-vafa} to determine quickly
what $\nn=1$ theory is dual to a given conical Calabi--Yau.

A similar story for AdS$_4$ has so far remained more mysterious,
although \cite{lee, lee-lee-park, kim-lee-lee-park} have predicted 
certain aspects of the superconformal theory with a
crystal model. One reason for this is that, whereas a
four--dimensional Yang--Mills coupling $g_{{\rm YM}\,4}$ runs
logarithmically, with a coefficient that can be tuned to zero with
a judicious choice of field contents, its analogue $g_{{\rm
YM}\,3}$ in three dimensions runs already classically. This is
related to the fact that the dilaton for a D2 brane solution is
not constant (nor is the near--horizon limit for its supergravity
solution of the form AdS$_4\times M_6$). Even though the M2
solution does not have these problems, not enough is understood of
its superconformal fixed point.

This situation has begun to change recently. Following the
discovery of a superconformal $\nn=8$ Chern--Simons theory
\cite{bagger-lambert1,
bagger-lambert2,bagger-lambert3,gustavsson}, a version of the
correspondence,
in the spirit of \cite{schwarz-chernsimons},
 has been proposed in which
the CFT has a Lagrangian description: it is a $\nn=6$
superconformal theory \cite{aharony-bergman-jafferis-maldacena}.
In a certain regime, its gravity dual is AdS$_4 \times \cc\pp^3$
in IIA. Other solutions dual to CFT's with a Lagrangian
descriptions have since been proposed: an orbifold of AdS$_4
\times \cc\pp^3$ in \cite{hosomichi-lee-lee-lee-park,
benna-klebanov-klose-smedback}, solutions that come from extrema
of the same $\nn=8$ four--dimensional effective theory in
\cite{benna-klebanov-klose-smedback,ahn,ahn2}, cases with orientifolds
\cite{armoni-naqvi,hosomichi-lee-lee-lee-park2,
aharony-bergman-jafferis}, and a squashed
$\cc\pp^3$ \cite{ooguri-park}. 
Earlier proposals for superconformal duals to $\nn=3$ AdS$_4$ have
been made in \cite{billo-fabbri-fre-merlatti-zaffaroni,
gukov-tong2, lee-yee, yee}.

In this note, we propose a series of $\nn=3$ quiver theories dual
to the near--horizon limit of M2 branes at certain hyperK\"ahler
singularities. Even if the engineering of our theories is a
straightforward variation on
\cite{aharony-bergman-jafferis-maldacena}, we believe it is
important to enlarge the class of duals available for study. We
regard our results as a first step towards obtaining as rich a
variety of AdS$_4$/CFT$_3$ duals as presently available for
AdS$_5$/CFT$_4$.

We describe the moduli spaces of these quiver theories as
hyperK\"ahler quotients, which allows us to show in an elegant way
their equivalence to the proposed dual backgrounds.

In section \ref{sec:theories} we will introduce our $\nn=3$
superconformal quiver theories.  In \ref{sec:geometry} we compute
their moduli space; while this has been done before
\cite{imamura-kimura}, we will point out that these manifolds are
hyperK\"ahler toric, so that their geometry is summarized neatly
by an array of two--dimensional vectors (see figure \ref{fig:fan}
below).
We also consider the orientifold of these theories.

In section \ref{sec:branes} we then engineer the theories in
string theory (again as done in \cite{imamura-kimura}); it turns
out that this can be done with a configuration of branes whose
inclinations are the same as the vectors in the hypertoric fan. We
use a duality chain \cite{gauntlett-gibbons-papadopoulos-townsend}
to explain that coincidence, and hence to derive infinitely many
instances of the AdS/CFT correspondence.
We also discuss the
reduction to IIA theory on AdS$_4 \times M_6$ in the large $K$
limit.


\section{The theories and their moduli space} 
\label{sec:theories}

We will consider $\nn=3$ Chern--Simons actions. Such theories are rigid, in the sense that the
Lagrangian is fixed by the gauge groups, Chern-Simons levels, and
matter representations. Moreover, such an $\nn=3$ theory can be
constructed given such data \cite{gaiotto-yin}. The gauge group is
${\rm U}(N_1) \times \ldots \times {\rm U}(N_n)$; there is also
matter, in the form  of $\nn=2$ multiplets $A_i$, $B_i$
(``chiral'', in the sense that they depend only on
$\theta\equiv\theta^1+ i \theta^2$), for $i=1,\ldots,n$.

\begin{figure}[ht]
\begin{picture}(200,120)(0,0)
    \put(150,0){\includegraphics[width=10em]{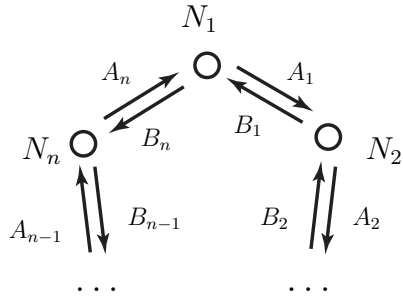}}
    \put(170,80){\scalebox{.8}{$A_n$}}
    \put(185,55){\scalebox{.8}{$B_n$}}
    \put(240,80){\scalebox{.8}{$A_1$}}
    \put(220,60){\scalebox{.8}{$B_1$}}
    \put(135,20){\scalebox{.8}{$A_{n-1}$}}
    \put(180,25){\scalebox{.8}{$B_{n-1}$}}
    \put(265,25){\scalebox{.8}{$A_2$}}
    \put(230,25){\scalebox{.8}{$B_2$}}
    \put(160,0){\scalebox{1.2}{$\ldots$}}
        \put(240,0){\scalebox{1.2}{$\ldots$}}
    \put(200,100){$N_1$}
    \put(140,50){$N_n$}
   \put(270,50){$N_2$}
    \end{picture}
\caption{\small The quiver.}
\label{fig:quiver}
\end{figure}

These fields transform
in the bifundamentals of the gauge group, in the way shown in
figure \ref{fig:quiver}.

The requirement that there be $\nn=3$ supersymmetry fixes the
action:
\begin{equation}
    \label{eq:S}
    S= \sum_i \Big[S_{{\cal N}=2 \, {\rm CS}}(k_i,V_i)+ \int d^4 \theta {\rm Tr}
    \Big(
    e^{-V_i} A_i^\dagger e^{V_{i+1}} A_i + e^{V_i} B_i e^{-V_{i+1}} B_i^\dagger \Big)+ \int d^2 \theta W \Big]\ ;
\end{equation}
the $\nn=2$ superpotential $W$ is
\begin{equation}
    \label{eq:W} W = \sum_{i=1}^n \frac 1 {k_i} Tr(B_i A_i - A_{i-1} B_{i-1})^2 \ .
\end{equation}
The levels $k_i$, $i=1,\ldots,n$ satisfy
\begin{equation}\label{eq:zerosum}
    \sum_{i=1}^n k_i =0\ .
\end{equation}
The $\nn=2$ Chern--Simons action was introduced in \cite{zupnik-pak,ivanov}, and shown to be conformal in
\cite{avdeev-kazakov-kondrashuk,gaiotto-yin}; the general
structure of the $\nn=3$ Chern--Simons theories was first developed in \cite{zupnik-khetselius, kao,
kao-lee-lee}.

We will engineer these actions from string theories in section \ref{sec:branes}
(as also done in \cite{imamura-kimura}).
For the time being, however, we will study their moduli spaces and show that
they suggest a gauge--gravity duality.

\subsection{The moduli space}
\label{sec:ms}

    We will first look at the moduli space in the case in which
    all the gauge group ranks are one: $N_i=1$, $i=1,\ldots,n$.

   By varying (\ref{eq:W}), one gets the F--terms\footnote{When one varies
   (\ref{eq:W}) with respect to $A_i$, one obtains (\ref{eq:f}) multiplied by
   $B_i$, and vice versa; this might raise the question of whether we are overlooking
   other branches, but one can show a posteriori that this is not the case.}
\begin{equation}\label{eq:f}
    (k_i + k_{i+1}) A_i B_i = k_i A_{i+1} B_{i+1} + k_{i+1} A_{i-1} B_{i-1}\ .
\end{equation}
    The D--term equation have a similar form, with $A_i B_i$ replaced by
    $|A_i|^2 - |B_i|^2$. One can then write D--term and F--term together as
\begin{equation}
    \label{eq:qq}
    (k_i + k_{i+1})\, q_i^\dagger \sigma_{\alpha} q_i = k_i \, q_{i+1}^\dagger \sigma_{\alpha} q_{i+1}
+ k_{i+1}\, q_{i-1}^\dagger \sigma_{\alpha} q_{i-1}\  ,
\end{equation}
where $q_i=(A_i, \bar B_i)^t$ and $\sigma_{\alpha}$ are the Pauli matrices.
\footnote{One can use quaternions already at the level of the action if one uses $\nn=1$
superfields; in the abelian case, the potential reads
$\int d^2 \theta_1 \sum_i \frac1k_i(q_i^\dagger \sigma_\alpha q_i-q_{i-1}^\dagger \sigma_\alpha q_{i-1})^2$.}
It is also convenient to write this as
\begin{equation}\label{eq:M}
    0= \sum_j M_{ij}\,q_j^\dagger \sigma_{\alpha} q_j\ , \qquad
    M=\left(\begin{array}{cccccc}
    k_1 + k_2 & -k_1 &0 &\cdots&0& -k_2 \\
    -k_3 & k_2 +k_3 & -k_2 &0&\cdots &0\\
    0 & -k_4 & \ddots & \ddots& \ddots & \vdots\\
    \vdots & \ddots &\ddots & \ddots &\ddots &\vdots\\
    0 & \cdots & 0 &-k_n & k_{n-1} + k_n & -k_{n-1}\\
    -k_n & 0 & \cdots & 0 & -k_1 & k_1 +k_n
    \end{array}\right) \ .
\end{equation}
    If one remembers that $\sum_i k_i=0$, it is easy to see that this
    matrix has rank $n-2$. Hence, at this point, we have found
    $3(n-2)$ independent real equations on $4n$ real coordinates.

We now turn to the gauge fields. One might think that the moduli
space obtained so far has to be quotiented by $U(1)^n$. However,
it is easy to see that the vector $\aa_+\equiv\sum_i \aa_i$ does not couple to
any scalar. It then turns out that one of the remaining $n-1$ gauge transformations
gets discretized by the Chern--Simons coupling $\sum_i k_i \aa_i F_i$.

Let us see this more in detail. The gauge transformations act as
\begin{equation}\label{eq:gaugetr}
    \aa_i \to \aa_i + d \lambda_i \ ; \qquad
    A_i \to e^{i (\lambda_i-\lambda_{i+1})} A_i \ , \qquad B_i \to e^{i(\lambda_{i+1}-\lambda_i)} B_i\ .
\end{equation}
It can be helpful to think of these transformations by going to a new basis of vectors:
\begin{equation}\label{eq:tildea}
    \aa_+ = \sum_{i=1}^n \aa_i \ , \qquad \tilde\aa_i = k_{i+1} \aa_i - k_i \aa_{i+1}\ , \qquad
    \aa_- = \sum_{i=1}^{n-1} \aa_i-\aa_n\ .
\end{equation}
$\aa_+$ does not act on the scalars; as for for the gauge
transformation of $\tilde\aa_i$, the $i$--th row of $M$ gives the
charges under it of the $j$--th scalar. Only $n-2$ of the
$\tilde\aa_i$ are independent (for the same reason that $M$ had
rank $n-2$), and so we complete the basis with $\aa_-$. Both
$\aa_\pm$ do not couple to themselves but to each other: in this
basis, there is a term $k_n \aa_+ d \aa_-$, but no $\aa_+ d \aa_+$
nor $\aa_- d \aa_-$. Since the choice of $\aa_-$ in
(\ref{eq:tildea}) might seem arbitrary, let us describe the
situation a bit more invariantly. We can think of $\int \sum_i k_i
\aa_i \wedge d \aa_i$ as a quadratic form ${\cal K}={\rm
diag}(k_1,\ldots, k_n)$ pairing the vector $(\aa_1,\ldots, \aa_n)$
to itself. Then $\aa_\pm$ are orthogonal to themselves because
they are in the kernel of ${\cal K}$ (thanks to $\sum_i k_i=0$);
and the $\tilde\aa_i$ in (\ref{eq:tildea}) are orthogonal to
$\aa_\pm$ for $i=2,\ldots, n-1$.

In any case, the constant gauge transformations generated by
$\aa_-$ do not quotient the moduli space. We can see this in the
same way as in \cite{mukhi, aharony-bergman-jafferis-maldacena},
with $\aa_\pm$ playing the role of $A_b$, $A_{\bar b}$ in that
reference. Since $\aa_+$ only appears through the coupling $k_n
\aa_+ d \aa_-$, we can dualize it to a periodic scalar $\tau$, and
add it to the configuration space. This scalar is charged under
$\aa_-$, and it can be used to fix its gauge transformation (up to
a discrete component that we will analyze shortly). This means
that only the gauge transformations of the $\tilde\aa_i$ should
act on the moduli space. Hence the matrix of charges can be taken
to be $M$; since this matrix has rank $n-2$, if one wants to avoid
redundancies, one can erase from it the first and last rows.

Let us now look at the possible residual gauge symmetry $\lambda_-$ from $\aa_-$. Looking at (\ref{eq:gaugetr}),
we see that it acts with charge 1 on $A_{n-1}$, with charge $-1$ on $A_n$, and 0 on the remaining
$A_i$. $\tau$ has period $2\pi$, but transforms as $\tau\to \tau+ k_n \lambda_-$ (just as in
\cite{aharony-bergman-jafferis-maldacena}). This results in a discrete action on $A_{n-1}$ and $A_n$.
One might, however, worry that this result depends on the particular choice of $\aa_-$ we have
made in (\ref{eq:tildea}). We can actually summarize both the discrete part of the gauge transformations
and the U$(1)^{n-2}$ by defining the group
\begin{equation}\label{eq:N}
    {\bf N}\equiv {\rm Ker}(\beta)\ ,
\end{equation}
where
\begin{equation}\label{eq:beta}
    \beta: {\rm U}(1)^n\to {\rm U}(1)^2 \qquad \beta= \left(\begin{array}{ccccc}
   1 & 1 & \ldots & 1 & 1\\
  k_1 & k_1 + k_2 & \ldots &k_1+\ldots + k_{n-1}& 0
    \end{array}\right)\ .
\end{equation}

Note that the ambiguity in defining $\beta$ by taking different
linear combinations of its rows results in a map with the same
kernel. Moreover, when we later relate the column of this matrix to
the fivebrane charges used to engineer the field theory, this
ambiguity is precisely the choice of an Sl$(2, \zz)$ duality
frame.

Let us check that ${\bf N}$ is the right residual gauge group.
An element $(e^{i \lambda_1}, \ldots, e^{i \lambda_n})\in {\bf N}$ has to satisfy
\begin{equation}\label{eq:Nexpl}
    e^{i \sum_i p_i \lambda_i}=1\ , \qquad e^{i \sum_i \lambda_i}=1 \ ,
\end{equation}
where
\begin{equation}\label{eq:kp}
    p_1=k_1\ , \qquad p_2=k_1+k_2\ , \ldots \qquad p_{n-1}= k_1 + \ldots +k_{n-1}, \qquad p_n=0
\end{equation}
are the entries on the first row of $\beta$, (\ref{eq:beta}).
There is a continuous component to its general solution: we can take $\theta_i = \sum_{j=2}^{n-1} M_{ji}\hat\theta_j$, since, as remarked above, the rows of $M$ generate the kernel (over $\rr$) of $\beta$.
The $\hat \theta$ span a subgroup ${\rm U}(1)^{n-2}\subset {\bf N}$, which corresponds to the gauge transformation
for the vectors $-k_{j-1}\aa_j + k_j \aa_{j+1}$.

Let us now look at the discrete part of ${\bf N}$. For example,
in the case considered in \cite{aharony-bergman-jafferis-maldacena},
$\beta$ is given by
\begin{equation}\label{eq:betaabjm}
    \left(\begin{array}{cc}
  1 & 1 \\
  0 & k
  \end{array}\right)\ .
\end{equation}
In this case, (\ref{eq:Nexpl}) gives
$e^{k i \lambda_2}=1$, $e^{i(\lambda_1 + \lambda_2)}=1$,
which is solved by $e^{i\lambda_1}=e^{-i\lambda_2}=\omega_k^j$,
where $\omega_k$ is a $k$--th root of unity and $j=1,\ldots, k$.
In this case ${\bf N}=\zz_k$ (there is clearly no continuous component)
and the moduli space is $\cc^2/\zz_k$.

In the general case, one can consider for example the action of the gauge transformation
corresponding to the vector $\aa_-$ above. This acts only on the last two hypermultiplets:
$\lambda_i=0$ but for $\lambda_{n-1}$ and $\lambda_n$. This is the residual discrete gauge invariance of $\aa_-$ mentioned earlier. Equation (\ref{eq:Nexpl}) now can be solved by $e^{i\lambda_{n-1}}=e^{-i\lambda_n}=\omega_{k_n}^j$ (with $\omega_{k_n}$ a $k_n$--th root of unity,
and $j=1,\ldots, k_n$).
This would seem a discrete subgroup of ${\bf N}$. However, when the $k_i$ are coprime, one can see that this discrete subgroup is actually
a subgroup of the U$(1)^{n-2}$ that we already considered. When the $k_i$ are not coprime, there is
a genuinely discrete component to ${\bf N}$, that we can take to be generated by
${\bf n}=(0,\ldots,0,\omega_K, \omega_K^{-1})$, where $K$ is
the l.c.d. of the $k_i$, and $\omega_K$ is once again a $K$--th root of unity.
We will actually see in section \ref{sec:branes} that the case in which the $k_i$
are not coprime has a particular relevance to us.

We have seen that, for $n>2$, ${\bf N}$ will necessarily have a continuous part.
One might wonder, though, whether one could get rid of some of the
coordinates and keep only the discrete part, so that the singularity
in the moduli space is still an orbifold. We will see in the next
subsection that this is typically not the case.

Summing up, the moduli space that we have obtained in this section is given by
\begin{equation}\label{eq:ms}
{\cal M}    = \frac{\{\ q_i=(A_i,\bar B_i)\in \cc^{2 \cdot n}\  |
\ \sum_j M_{ij}q_j^\dagger \sigma_{\alpha} q_j=0\ \}}
{(A_i,B_i)\sim ({\bf n}_i A_i , {\bf n}_i^{-1} B_i)\, ,\ {\bf
n}\in {\bf N}}
\end{equation}
where $M$ has been defined in (\ref{eq:M}) and ${\bf N}$ has been defined in (\ref{eq:N}),(\ref{eq:beta}).
Notice that the numerator consists, as mentioned earlier, of $3(n-2)$ real equations; and that
the denominator is a group whose continuous part is $(n-2)$--dimensional. Hence one expects a
dimension
\begin{equation}
    4n - 3(n-2) -(n-2) = 8 \ .
\end{equation}
In fact, we will now see that this quotient has already been studied by mathematicians,
and that one can say a great deal more about ${\cal M}$.

\subsection{Geometrical interpretation}
\label{sec:geometry}

Let us first take a step back. An easy way to produce a \ka\ manifold of complex dimension
$k-j$ is the so--called \ka\ quotient construction. One starts from $\cc^k$, with
${\rm U}(1)^j$ acting on it. Using the \ka\ form as a symplectic form,
one can compute a Hamiltonian $\mu$ for each of the $j$ U$(1)$ actions; this $j$--uple of
functions is called the moment map for the
${\rm U}(1)^j$ action. Since the latter preserves the Hamiltonian,
it acts in particular on each of the level sets $\mu=\mu_0$. So if one considers
\begin{equation}\label{eq:mu}
    \cc^k//(\cc^*)^j\equiv\frac{\{\mu=\mu_0\}}{{\rm U}(1)^j}\ ,
\end{equation}
one ends up with a manifold of complex dimension $k-j$.
 One can then ``pull back''
the \ka\ structure from $\cc^k$ to the quotient (\ref{eq:mu}), which is then a \ka\ manifold
itself.

There is a similar procedure for hyperK\"ahler manifolds. One starts now from $\hh^k$,
again with a ${\rm U}(1)^j$ action.
One now has an ${\rm SU}(2)$ triplet of moment maps $\vec{\mu}$ (a ``hyperK\"ahler moment map''
\cite{hitchin-karlhede-lindstrom-rocek,goto,bielawski-dancer,gauntlett-gibbons-papadopoulos-townsend}).
One can then quotient by ${\rm U}(1)^j$:
\begin{equation}\label{eq:hypermu}
    \frac{\{\vec{\mu}=\vec{\mu_0}\}}{{\rm U}(1)^j}\ ,
\end{equation}
similarly to (\ref{eq:mu}).  This time, one has lost $3j+j$ real coordinates, or $j$
quaternionic coordinates. It turns out that the manifold defined in this way can be
given a hyper\ka\ structure.

Let us see the quotient (\ref{eq:hypermu}) more in detail, following in particular
\cite{bielawski-dancer}. First of all, let us think of our $q_i = (A_i, \bar B_i)$
as the $i$--th quaternionic coordinate on $\hh^n$. We can act with a natural ${\rm U}(1)$ on
each of these coordinates: the $i$--th such action reads
\begin{equation}\label{eq:ithact}
    (A_i, \bar B_i) \to (e^{i\theta} A_i, e^{i \theta}\bar B_i)\ .
\end{equation}
The hyperK\"ahler moment
map that generates an action is the triplet
\begin{equation}
    q^\dagger_i \sigma_\alpha q_i \ .
\end{equation}
If one has many ${\rm U}(1)$ actions acting on the $q_i$ according to a charge matrix $M$,
one ends up with a hyperK\"ahler moment map $M_{ij} q^\dagger_i \sigma_\alpha q_i $, which
is exactly what enters in (\ref{eq:ms}).

One can think of each of these moment
maps as a different Hamiltonian for the U$(1)$ actions.
It is then not a surprise that the U$(1)$ actions leave the
value of $M_{ij} q^\dagger_i \sigma_\alpha q_i $ invariant. In particular, the locus
in the numerator of (\ref{eq:ms}) is left invariant by the action in the denominator.
Hence we see that (\ref{eq:ms}) is in fact a particular case of the general definition
of hyperK\"ahler quotient (\ref{eq:hypermu}) \cite{bielawski-dancer}.

We have actually glossed over the fact that our action is not exactly an U$(1)^{n-2}$ action,
but it might have a discrete component: the abelian group in the denominator of (\ref{eq:ms})
was defined in (\ref{eq:N}). Fortunately, the definition in \cite{bielawski-dancer} considers
exactly the same type of abelian action, and hence we can use their results.

To say more, let us introduce some more notation. Call $u_i \in \rr^2$ the columns of (\ref{eq:beta}).
(These vectors are
the analogue of the fan in toric geometry.) In terms of the $p_i$ defined in
(\ref{eq:kp}), they look like figure \ref{fig:fan}:

\begin{figure}[ht]
\begin{picture}(200,150)(0,0)
    \put(120,0){\includegraphics[width=25em]{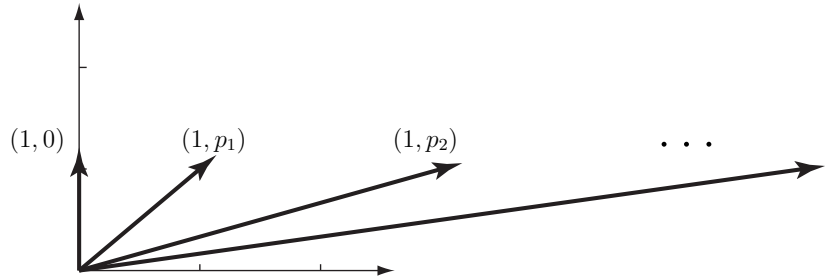}}
    \put(105,70){\scalebox{.8}{$(1,0)$}}
    \put(170,70){\scalebox{.8}{$(1,p_1)$}}
    \put(250,70){\scalebox{.8}{$(1,p_2)$}}
    \put(350,70){\scalebox{1.6}{$\ldots$}}
    \end{picture}
\caption{\small The vectors $u_i$ of the ``hypertoric fan''.}
    \label{fig:fan}
\end{figure}

We have remarked already that in our case ${\cal M}$ is a cone. At
the tip of this cone one will find a singularity, but one would
like the base of the cone to be nonsingular.
 Theorem 4.1 in
\cite{bielawski-dancer} guarantees that this is the case for our
$\beta$, as long as $k_i \neq 0$ for all $i=1,\ldots,n$. As for
the conical singularity, Theorem 3.3 in \cite{bielawski-dancer}
gives a criterion to decide when it is an abelian orbifold: in our
case, the criterion says that the singularity is an abelian
orbifold if and only if there are only two different $u_i$ (the
columns of $\beta$). This is in agreement with what we know about
the $\nn=6$ theory in \cite{aharony-bergman-jafferis-maldacena}
(for which $n=2$, and hence there are indeed only two $u_i$, namely
the two columns of (\ref{eq:betaabjm})), and
for the theory in \cite{benna-klebanov-klose-smedback} (in which
$k_{\rm odd}=k$ and $k_{\rm even}=-k$, so that $u_{\rm
odd}={1\choose k}$ and $u_{\rm even}={1\choose 0}$).

The ``base''  (or horizon) $B_7$ of the cone,
\begin{equation}
    {\cal M} \equiv {\rm Cone}(B_7)\ ,
\end{equation}
is called 3--Sasakian whenever the cone ${\cal M}$ is hyperK\"ahler (which is our case).
The $B_7$ we are getting in this paper were defined and studied by Boyer, Galicki and Mann \cite{boyer-galicki-mann1,boyer-galicki-mann2}, as part of a more general construction; a
review of the part relevant to us is given in \cite[Section 7]{boyer-galicki}.

\subsection{Comparison with four--dimensional quiver theory}
\label{sec:4d}

In this paper we are using quivers to define $\nn=3$ theories in three dimensions.
Quivers have been used for a long time to define $\nn=1$ theories in four dimensions,
and it is natural to wonder if there is any relation between the moduli spaces in the
two cases. In this subsection, we will call the moduli space defined in (\ref{eq:ms})
(the one relevant for this paper) ${\cal M}_{\rm d=3}$. Likewise, we will call ${\cal M}_{\rm d=4}$
the moduli space for the $d=4$, $\nn=1$ theory defined from the same quiver and superpotential.

At first blush, one might think that there should be little relation between the two.
For example, for the $d=3$, $\nn=6$ theory in \cite{aharony-bergman-jafferis-maldacena},
${\cal M}_{\rm d=3}=\cc^4/\zz_k$; the $d=4$, $\nn=1$ theory with the same quiver and superpotential
 has ${\cal M}_{\rm d=4}=$ the conifold \cite{klebanov-witten}.
These two spaces appear to be very different.

The difference between ${\cal M}_{\rm d=3}$ and ${\cal M}_{\rm d=4}$ can be traced to
the D--term contribution to the potential. In four dimensions, the D--term
equation for our quiver would have read
\begin{equation}\label{eq:d4d}
    |A_i|^2 - |B_i|^2- |A_{i-1}|^2 + |B_{i-1}|^2= \mu_i \ \qquad {\rm (in\ four\ dimensions)}
\end{equation}
with $\mu_i$ a Fayet--Iliopoulos parameter. In our case, the D--term component ($\alpha=3$)
of (\ref{eq:M}) can be written as
\begin{equation}
    \frac{|A_i|^2 - |B_i|^2- |A_{i-1}|^2 + |B_{i-1}|^2}{k_i}=
    \frac{|A_{i+1}|^2 - |B_{i+1}|^2- |A_i|^2 + |B_i|^2}{k_{i+1}} \ .
\end{equation}
We can rewrite this as
\begin{equation}\label{eq:d3d}
    |A_i|^2 - |B_i|^2- |A_{i-1}|^2 + |B_{i-1}|^2= \lambda \, k_i\ , \qquad  \forall i\ ,
\end{equation}
for some $\lambda$. (\ref{eq:d3d}) is now formally identical to (\ref{eq:d4d}).
The difference is that $\lambda$ is an arbitrary parameter, whereas the $\mu_i$
in (\ref{eq:d4d}) have to be fixed to some value (and are moduli for the geometry).

Another, related, difference between the two theories arises in the action by the gauge group.
In the four--dimensional theory, one mods out by U$(1)^{n-1}$; in the three--dimensional
theory, by U$(1)^{n-2}$.

Putting these two remarks together, we can say that
\begin{equation}\label{eq:43}
    {\cal M}_{\rm d=4}={\cal M}_{\rm d=3}//\cc^*
\end{equation}
or, in other words, that ${\cal M}_{\rm d=3}/{\rm U}(1)$ is
foliated by copies of ${\cal M}_{\rm d=4}$ obtained for different values
of the four--dimensional FI parameters $\mu_i=\lambda k_i$.
(The leaves will change topology at some special value of the $\lambda$.)

This relation to the moduli spaces of the $3+1$ Yang-Mills theory
with the same quiver persists when the Chern--Simons--matter theory
only possesses ${\cal N}=2$ supersymmetry. In that case, the
D--terms are the same as in (\ref{eq:d3d}), but the F--term
equations are different.

It is also interesting to consider
turning on (generically ${\cal N}=2$) FI parameters in the
Chern-Simons theory. The quaternionic equations \ref{eq:qq} are modified to
\begin{equation}
    \label{eq:qqFI}
    \frac{1}{k_{i+1}} (q_i^\dagger \sigma_{\alpha} q_i -q_{i+1}^\dagger \sigma_{\alpha} q_{i+1}) - \zeta_{\alpha,\,i+1} = \frac{1}
    {k_i} (q_{i-1}^\dagger \sigma_{\alpha} q_{i-1}\ - q_i^\dagger \sigma_{\alpha} q_i) - \zeta_{\alpha,\,i} \  .
\end{equation}
The $\alpha=3$ component of (\ref{eq:qqFI}) is again implied by
(\ref{eq:d4d})
with $\mu_i = (\zeta_{3,\,i} + \lambda) k_i$, so again the moduli space
of the $2+1$ theory is a K\"ahler quotient of the $3+1$ moduli
space. (Nonzero $\zeta_{1\,i}$ and $\zeta_{2\,i}$ would come from a modification
of the $\nn=2$ superpotential (\ref{eq:W}).)

For example, in the case of the quiver with two nodes, used in three dimensions
in \cite{aharony-bergman-jafferis-maldacena}, and in four dimensions by
\cite{klebanov-witten}, the statement (\ref{eq:43}) is, for $k=1$,
\begin{equation}
    \cc^4//\cc^*= {\rm conifold}\ :
\end{equation}
if one takes $\cc^4/{\rm U}(1)$ and one fixes $\lambda$, one obtains a copy of
the resolved conifold for $\lambda\neq0$, and of the singular conifold for
$\lambda=0$; by varying $\lambda$, one sweeps the entire $\cc^4/{\rm U}(1)$.

The quivers in figure \ref{fig:quiver} were used for four--dimensional $\nn=1$ theories
in \cite{gubser-nekrasov-shatashvili}. Their description of the theory is in terms
of two ``big matrices'' $X_1, X_2$ whose blocks summarize our $A_i$ and $B_i$ respectively,
and of a third, block--diagonal, matrix $\Phi$. Their superpotential is
\begin{equation}
    W= {\rm Tr}\Big(\Phi[X_1,X_2]+ M \Phi^2\Big)    \ ,
\end{equation}
where $M$ is also block--diagonal; if one integrates out $\Phi$, one obtains our superpotential
(\ref{eq:W}), with their $m_i$ equal to our $k_i$.
The moduli spaces for \cite{gubser-nekrasov-shatashvili} are
\begin{equation}\label{eq:gencon}
    {\cal M}_{\rm d=4}= \{\ (u,v,z,w)\in \cc^4 \ | \ uv = \Pi_{i=1}^n (z- k_i w)\ \}\ ,
\end{equation}
which they call ``generalized conifolds''. We can then give a description of
our ${\cal M}_{\rm d=3}$ defined in (\ref{eq:M}), by reading (\ref{eq:43}) backwards.

\subsection{An example}
\label{sec:example}

We will now use \cite{boyer-galicki} to gain insight on perhaps
the simplest new set of examples discussed in this paper: the case
in which the number of nodes $n=3$. As we pointed out, we know
from \cite[Theorem 3.3]{bielawski-dancer} that in this case the
conical singularity is not an abelian orbifold. In this case, $M$
has only one row (repeated three times):
\begin{equation}
 ( k_3 \ k_1 \ k_2 )
\end{equation}
which is the ``charge vector'': its $i$--th entry tells us the charge of
$(A_i,\bar B_i)$ under the only ${\rm U}(1)$. There is then only one triplet
of equations, and only one quotient\footnote{For simplicity we assume here
that the $k_i$ are coprime. The case in which they are not is a discrete quotient
of the case in which they are, as in (\ref{eq:zk}).}:
\begin{equation}\label{eq:M3}
    {\cal M}_{n=3}=\frac{\{\ (A_i,B_i)\ |\
    \sum_i k_{i-1} A_i B_i=0\ , \sum_i k_{i-1}(|A_i|^2-|B_i|^2)=0 \ \}}
    {(A_i,B_i)\to (e^{i k_{i-1}\theta} A_i, e^{-i k_{i-1}\theta} B_i)}\ .
\end{equation}
This manifold is a cone. One would be tempted to fix the radial gauge by imposing (say) the further
constraint $\sum_i k_{i-1}(|A_i|^2+|B_i|^2)=2$.
Since $\sum_i k_i=0$, the $k_i$ cannot all be positive, and one would end up with hyperboloids that
are not particularly illuminating. One can, however, change variables and obtain clearer equations.
Suppose for example $k_{1,2}>0$, $k_3 <0$. If one defines
\begin{equation}\label{eq:kprime}
    k_3'=-k_3 \ , \qquad k_{1,2}'=k_{1,2}\ ; \qquad A_3'=-B_3\ , \qquad B_3'=A_3\ ,
\end{equation}
equation (\ref{eq:M3}) is invariant in form:
\begin{equation}\label{eq:M3p}
    {\cal M}_{n=3}=\frac{\{\ (A'_i,B'_i)\ |\
    \sum_i k'_{i-1} A'_i B'_i=0\ , \sum_i k'_{i-1}(|A_i'|^2-|B_i'|^2)=0 \ \}}
    {(A_i',B_i')\to (e^{i k'_{i-1}\theta} A'_i, e^{-i k'_{i-1}\theta} B'_i)}\ ;
\end{equation}
with the difference that now
\begin{equation}
    k'_i >0 \ .
\end{equation}

This change of variables is equivalent to flipping the sign of one
vector in the hypertoric fan. Such an operator does not alter the
resulting hyperK\"ahler geometry, which can be understood in terms
on the $(p,q)$ 5-branes we shall soon discuss, as the fact that
the angle at which an anti-fivebrane must lie to preserve $\nn=3$
supersymmetry is exactly the opposite of the corresponding
fivebrane; hence they are the same object.

We can now study the base of the cone by intersecting it with the further constraint
$\sum_i k'_{i-1}(|A'_i|^2+|B'_i|^2)=2$. One obtains
\begin{equation}\label{eq:un}
    \sum_i k_{i-1}' |A'_i|^2=\sum_i k_{i-1}' |B'_i|^2= 1\ , \qquad  \sum_i k'_{i-1} A'_i B'_i=0\ .
\end{equation}
If we define $x_i= \sqrt{k'_{i-1}} A_i'$, $y_i = \sqrt{k'_{i-1}} \bar B_i'$, equation (\ref{eq:un})
is equivalent to the statement that $x$ and $y$ can be used as the first two columns of a
U$(3)$ matrix $U$. There is a U$(1)$ worth of vectors $\in \cc^3$ that can be used a third column
of $U$. Hence, solutions to (\ref{eq:un}) are given by the quotient U$(3)/{\rm U}(1)$. Remembering
now the action on the quotient of (\ref{eq:M3p}), and reinstating the conical direction, we conclude that
\begin{equation}\label{eq:biq}
    {\cal M}_{n=3} = {\rm Cone}(B_7)\ , \qquad
    B_7\equiv{\rm U}(1) \backslash {\rm U}(3)/{\rm U}(1)\ .
\end{equation}
The left action is given by ${\rm diag}(e^{i k'_3 \theta},e^{i k'_1 \theta},e^{i k'_2 \theta})$,
and the right action is given by ${\rm diag}(1,1,e^{i\phi})$. Notice that, since each of the
${\rm U}(1)$ has determinant not equal to one, one cannot rewrite (\ref{eq:biq}) as
${\rm SU}(3)/{\rm U}(1)$ nor as ${\rm U}(1) \backslash {\rm SU}(3)$.

The base $B_7$ of the cone ${\cal M}_{n=3}$ in (\ref{eq:biq})
is sometimes called ``Eschenburg space'' \cite{eschenburg1};
it was later found in \cite{boyer-galicki} (where it was called
${\cal S}(k_3',k_2',k_1')$) that it is indeed 3--Sasakian -- or in other words, that
the cone ${\cal M}_{n=3}$ over $B_7$ is indeed hyper\ka.
$B_7$ in (\ref{eq:biq}) is a close relative to the Aloff--Wallach spaces $N(k,l)$ considered
in the early Kaluza--Klein literature \cite{castellani-romans,duff-nilsson-pope,
sorokin-tkach-volkov-11-10};
those spaces are $\nn=3$ only for $k=l=1$ (and $\nn=1$ otherwise), whereas
(\ref{eq:biq}) is $\nn=3$ for any $k'_i$. Only $N(1,1)$ is a particular case of
Eschenburg space, but it would require $k_1=k_2=k_3'=1$, which does not occur for us.

Various aspects of the topology and geometry of (\ref{eq:biq})
have been studied in
\cite{eschenburg1,eschenburg2,boyer-galicki,chinburg-escher-ziller,lee-yee},
where in particular it is shown that its non--zero cohomology
groups are $H^2(B_7,\zz)=H^5(B_7,\zz)=\zz$, $H^4(B_7,\zz)=\zz_{k_1
k_2 + k_1 k_3'+ k_2 k_3'}$. \cite{lee-yee, yee} also considered
the problem of finding their field theory dual.

The isometry group of (\ref{eq:biq}) is ${\rm U}(2)\times {\rm U}(1)$. In the field theory,
an ${\rm SU}(2)$ subgroup of this appears as an R--symmetry;
a U$(1)$ is the symmetry generated by the current $\ast {\cal
F}_+$ (which has now become a total symmetry); a further U$(1)$ is
due to the symmetry $(A_i,B_i) \to (e^{i\theta} A_i,e^{-i\theta}
B_i)$.

\subsection{More general quivers?}

The techniques introduced (or reviewed) so far in this section can be actually applied to
more general $\nn=3$ theories. Even though the string theory interpretation is as yet
problematic, as we will explain later, the moduli spaces can be computed
just as easily as for the quiver in figure \ref{fig:quiver}.

Let us consider a more general quiver (with bifundamental matter). In this
subsection, the index $a$ will run over the nodes and the index $i$ over the
hypermultiplets. The scalars in each hypermultiplet $q_i$ again make up
a complex doublet $(A_i,\bar B_i)$. We encode the quiver
in a charge matrix $X_{ai}$ that has value 1 if $A_i$ leaves the node $a$,
$-1$ if it exits from it, and 0 otherwise.
The F-- and D--terms then read
\begin{equation}\label{eq:Mgen}
    M^{\rm gen}_{ij}q_j^\dagger \sigma_{\alpha} q_j =0 \ , \qquad M^{\rm gen}\equiv
    \sum_a \frac1k_a X_{ai}X_{aj}.
\end{equation}
The U$(1)$ actions turn out to involve the same matrix, and one obtains
once again a hyper\ka\ quotient. Let us work it out.

The matrix $M^{\rm gen}$ in (\ref{eq:Mgen}) has a zero eigenvalue for any
(independent) loop in the quiver. For generic $k_a$, then, the rank of the
matrix is $\# ({\rm edges})-\# ({\rm loops})$.
Since the number of variables
is $\# ({\rm edges})$, it follows that the dimension is
\begin{equation}
    {\rm dimension}= 4 \,\# ({\rm loops})\ ;
\end{equation}
hence, for string theory applications, one would need a quiver
with two loops. The hypertoric fan is easy to work out. Each of
the two rows can be computed from one of the two loops. If one
chooses an overall orientation for the loop, the $i$--th entry of
the row is 1 if the field $A_i$ follows the orientation, $-1$ if
it runs opposite to it, and 0 if the edge does not belong to the
loop. The $\nn=3$ theory considered in
\cite{billo-fabbri-fre-merlatti-zaffaroni} is of this type; it has
$N(1,1)={\rm SU}(3)/{\rm U}(1)$ as an abelian moduli space.
(\cite{gukov-tong2} also considered a similar theory, although
only in the abelian case.) At the end of subsection
\ref{sec:example} we commented on how $N(1,1)$ is different from
(\ref{eq:biq}).

There is another possibility to consider. If one also imposes
\begin{equation}
    \sum_a k_a =0\ ,
\end{equation}
the rank of the matrix
$M^{\rm gen}$ in (\ref{eq:Mgen}) drops further by 1, and one obtains that
\begin{equation}
    {\rm dimension}= 4\,\Big(\# ({\rm loops})+1\Big)\ .
\end{equation}
The theories (\ref{eq:S}) are of this type. The matrix $M^{\rm
gen}$ is in that case essentially $M$ in (\ref{eq:M}) (except
that, in (\ref{eq:M}), we have multiplied the $i$--th row by
$k_i+k_{i+1}$, for better readability). The fan can again be easily
worked out. One of the rows corresponds to the single loop in the
quiver. The other row can be obtained by considering a path
connecting all nodes, and it depends on the levels $k_a$ in a way
similar to the one in (\ref{eq:beta}). To summarize, the abelian
moduli space of a general quiver $\nn=3$ Chern--Simons is still a
hyper\ka\ quotient. The reason we are only sketching these general
rules is that the use of these quivers in string theory is not yet
entirely clear. There are clear candidates to engineer these
theories, suggested by the hypertoric fan along the lines
described in the next section. However, checking that the
engineering of the theories actually works would involve
understanding the general rules for extracting an effective theory for
D3's intersecting general $(p,q)$--fivebranes, which are not at
the moment entirely clear.


\subsection{Orientifold theories and their moduli space}
\label{sec:orientifold}

In this section we will consider orientifolding the theories
introduced above. Later this will be related to a brane
construction with an O3 plane wrapped around the circle with the
D3 branes. We begin with an ${\cal N}=3$ quiver Chern-Simons
theory of the type defined by (\ref{eq:S}), such that the number
of nodes is even, and the levels are given by $2 k_i$. The reason
for these restrictions will be seen shortly. This generalizes the
orientifolds considered in \cite{hosomichi-lee-lee-lee-park2,
aharony-bergman-jafferis}. Here there is no enhanced supersymmetry
beyond ${\cal N}=3$ to begin with, hence the orientifolded theory
preserves the same supersymmetries.

The action of the orientifold on the hypermultiplets, $C_i = (A_i,
\ B_i^\dag)$, is given by
\begin{equation}
\begin{array}{c}\vspace{.2cm}
C_{A \ 2i+1} \rightarrow - M_{A B}\, C_{B \ 2i+1}^* \,J \ ,\\
C_{A \ 2i} \rightarrow -M_{A B}\, J\, C_{B \ 2i}^*\ ,
\end{array}
\end{equation}
where $M_{A B} = i\sigma_2$, and $J$ is the invariant
anti--symmetric matrix of the USp theory. It is easy to check
that this is a discrete symmetry (for even ranks) of the action
(\ref{eq:S}), which we now gauge.

The gauge groups become, alternatively, O$(2N_{2i+1})$ and USp$(2 N_{2i})$,
and the projected matter fields are naturally thought of
as hypermultiplets obeying a reality condition. As shown in \cite{aharony-bergman-jafferis}
the levels become $2 k_{2i+1}$ and $k_{2i}$, which is why the original levels were
chosen to be even. To be more explicit, we can take the
identification on the matter fields to be $B_{2i+1} = J\,
A_{2i+1}^T$ and $B_{2i} = A_{2i}^T\, J$. Therefore the $\nn = 2$
superpotential is given by
\begin{equation}
    \label{eq:WO}
    W = \sum_{i=1}^n \Big[\frac{1} {2 k_{2i-1}} {\rm Tr}(J A_{2i-1}^T A_{2i-1} - A_{2i-2} A_{2i-2}^T J)^2 +
    \frac{1}{2 k_{2i}} {\rm Tr}(A_{2i}^T J A_{2i} - A_{2i-1} J A_{2i-1}^T)^2 \Big]\ .
\end{equation}

The moduli space of a single whole M2 brane in this background
corresponds to taking all of the ranks equal to two, so that the
gauge groups are O$(2) = {\rm U}(1) \rtimes \zz_2$ and USp$(2) =
{\rm SU}(2)$. Generalizing the analysis of \cite{aharony-bergman-jafferis},
we will choose $A_i = x_i I + y_i J$. Now all of the matrices appearing
in the F-term and D-term equations commute, since $A_i^T = x_i -
y_i J$. To relate the resulting constraints to the hyperK\"ahler
quotient described above, it is convenient to change variables to
$u_i = x_i + i y_i$ and $v_i = x_i - i y_i$, so that the
combination that appears in the F-terms, $A_i^T A_i = J u_i
v_i$, and in the D-terms, $A_{2i-1}^\dag A_{2i-1} - A_{2i-1}^T
A_{2i-1}^* = J (|u_{2i-1}|^2-|v_{2i-1}|^2)$, $A_{2i}^\dag A_{2i}
+ J A_{2i}^T A_{2i}^* J = J (|u_{2i}|^2-|v_{2i}|^2)$.

Therefore, defining $q_i = (u_i, \ v_i^*)$, we see that the moduli
space is described by equations identical to \ref{eq:qq}. The
gauge symmetry is broken to U$(1)^{2n}$, given by the SO$(2)$ and
the U$(1)$ subgroup of USp$(2)$ generated by $i\sigma_2$.
Therefore we can apply the same reasoning as before to obtain a
hyperK\"ahler quotient.

There is an additional discrete quotient obtained as in
\cite{aharony-bergman-jafferis} from the $\zz_2$ component of the
O$(2)$ gauge groups, given by the element $\sigma_1$. In general,
acting by this transformation will not preserve the ansatz above,
but a simultaneous action in all $O(2)$ factors, together with a
particular $USp(2)$ transformation does act on the moduli space.
Its action is $A_{2j} \rightarrow i \sigma_3 A_{2j} \sigma_1$ and
$A_{2j+1} \rightarrow \sigma_1 A_{2j+1} i \sigma_3$ hence $q_i
=(u_i, \ v_i) \rightarrow (-v_i, \ u_i)$. Note that applying this
transformation twice is just $q_i \rightarrow - q_i$, which is
already quotiented, since the least common divisor of the levels is at
least two, and this transformation is present in the discrete
component of the hyperK\"ahler quotient.

\section{String theory interpretation}
\label{sec:branes}

We have defined in (\ref{eq:S}) a certain set of $\nn=3$ theories, and we have shown
that their moduli space ${\cal M}$ (defined in (\ref{eq:ms}))
is a (toric) hyperK\"ahler manifold of dimension 8.
This is a cone over a 3--Sasakian manifold $B_7$.

This fact suggests that these theories might actually be dual to AdS$_4 \times B_7$.
We will now show that this is indeed the case. The argument is very similar to
the one in \cite{aharony-bergman-jafferis-maldacena}.

\begin{figure}[ht]
\begin{picture}(200,150)(0,0)
    \put(150,0){\includegraphics[width=15em]{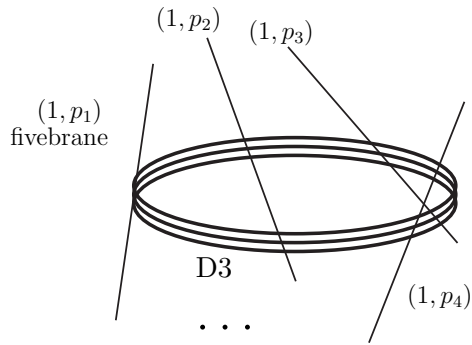}}
    \put(140,100){\scalebox{.8}{$(1,p_1)$}}
    \put(130,90){\scalebox{.8}{fivebrane}}
    \put(185,135){\scalebox{.8}{$(1,p_2)$}}
    \put(220,130){\scalebox{.8}{$(1,p_3)$}}
    \put(280,30){\scalebox{.8}{$(1,p_4)$}}
    \put(200,20){\scalebox{1.6}{$\ldots$}}
    \put(200,40){\small D3}
    \end{picture}
\caption{\small A cartoon of the brane configuration.}
\label{fig:branes}
\end{figure}

We start with a brane configuration in IIB. As shown in figure
\ref{fig:branes}, it consists of a D3 brane along directions
$0126$ (where $x^6$ is actually compactified to a circle), and of
several fivebranes. The $i$--th of these fivebranes has charges
$(1,p_i)$, and is extended along $012[37]_{\theta_i}
[48]_{\theta_i} [59]_{\theta_i}$, where $\tan(\theta_i)=p_i$. This
configuration preserves $\nn=3$ supersymmetry
\cite{kitao-ohta-ohta, bergman-hanany-karch-kol}.

Using the same reasoning as in \cite{aharony-bergman-jafferis-maldacena},
one can see that this configuration is such that it engineers the theory (\ref{eq:S}),
after relating the $p_i$ to the $k_i$ as in (\ref{eq:kp}). In fact, we can remark
already now that the $u_i$ of
the ``hypertoric fan'' in figure \ref{fig:fan} have coordinates exactly equal to
the $(p,q)$ charges of the fivebranes in figure 
\ref{fig:branes}.\footnote{This is reminiscent of the way some 
of the theories in \cite{gukov-tong2} were engineered.} This coincidence
will be explained shortly.

If one takes the figure literally, so that the D3 all fill up the entire $x^6$ circle,
the configuration actually engineers only the case in which all the ranks are equal,
$N_i = N_{\rm D3}$, $i=1,\ldots,n$. One could allow for D3 suspended between the fivebranes,
but for simplicity we will not do that, and in what follows we will consider $N_i=N$.

One can now T--dualize this configuration along direction 6 and then lift it to M--theory,
again similarly
to \cite{aharony-bergman-jafferis-maldacena}. For the time being, we forget
about the D3 branes; we will reintroduce them later.
The details of the duality chain
are explained in \cite{gauntlett-gibbons-papadopoulos-townsend}.
Jumping at the result,
the metric that one gets in M--theory is of the form
\begin{equation}\label{eq:hypertnut}
    ds^2= ds^2_{\rm Mink_3}+ U_{ab} d {\bf x}^a \cdot d {\bf x}^b + U^{ab}
    (d \phi_a + A_a)(d \phi_b+ A_b)\ ,
\end{equation}
where $a,b=1,2$. The internal eight--dimensional metric is a $T^2$
fibration over $\rr^6$ that generalizes to higher dimension the
Taub--NUT metric. The equations on the two--by--two matrix $U$ and
on $A$ for a general metric of the form (\ref{eq:hypertnut}) to be
hyperK\"ahler are also similar to the Taub--NUT case. The
particular solution one gets in our case is given by
\cite{bielawski-dancer,gauntlett-gibbons-papadopoulos-townsend}
\begin{equation}\label{eq:Mth}
    U_{ab} = U^{\infty}_{ab}+ \sum_{i=1}^n \frac{u_{ia} u_{ib}}{|\vec{x}+ p_i \vec{y}|}\
\end{equation}
where $\vec{x},\vec{y}$ are coordinates on $\rr^6$ (namely, directions $3,4,5$ and
$7,8,9$ respectively);
 recall that for us the $u_i={1\choose p_i}$ are the columns of
$\beta$. Notice also that
\begin{equation}
    \vec{x}+ p_i \vec{y}=q_i^\dagger \vec{\sigma} q_i
\end{equation}
is the general solution to (\ref{eq:M}), as becomes apparent by looking
at (\ref{eq:beta}) and (\ref{eq:kp}), whose rows generate the kernel of $M$.

One can think of the $T^2$ in (\ref{eq:hypertnut}) as follows. We started our
discussion of hyper\ka\ quotients in section \ref{sec:geometry} by considering the
$n$ U$(1)$ actions (\ref{eq:ithact}). Then we quotiented by a subgroup ${\bf N}$
(defined in (\ref{eq:N}) of that ${\rm U}(1)^n$. Since the matrix $M$ has for us
rank $n-2$, ${\bf N}$ has a continuous part U$(1)^{n-2}$, which leaves us with
an unquotiented U$(1)^2= T^2$. This is the $T^2$ in (\ref{eq:hypertnut}). In fact,
by looking at the degeneration loci
\begin{equation}\label{eq:deg}
    |\vec{x}+ p_i \vec{y}|=0 \ ,\quad i=1,\ldots,n\ ,
\end{equation}
we can conclude that $\phi^2=x^{10}$ (the M--theory circle), and that $\phi^1=
\tilde x^6$, the T--dual of the original $x^6$ direction in the IIB configuration
(the compact direction in figure \ref{fig:branes}).

Now, near each of the loci (\ref{eq:deg}), the
metric (\ref{eq:Mth}) in M--theory is non--singular: it behaves like a single Taub--NUT
solution. One expects, however, a singularity at the origin, where all
those loci coincide. If we focus on this singularity, we can drop the term $U_\infty$
from (\ref{eq:Mth}):
\begin{equation}
    U_\infty=0\ .
\end{equation}
We can now recognize that the metric found in \cite[Theorem
9.1]{bielawski-dancer} as being the one induced on our ${\cal M}$
by the hyperK\"ahler quotient procedure is the same as
(\ref{eq:hypertnut}).
Therefore we have shown that the moduli
space of the conformal field theory is exactly the geometry
obtained by dualizing and lifting the fivebranes to M--theory.

The last step is to reinstate the D3 branes. When we do include them,
we end up with a stack of M2 branes
at the tip of ${\cal M}$. In the near--horizon limit for these M2 branes, one ends up with
\begin{equation}\label{eq:ads}
    {\rm AdS}_4 \times B_7
\end{equation}
(recall that ${\rm Cone}(B_7)= {\cal M}$), with Freund--Rubin fluxes. Since $B_7$ is a
3--Sasakian manifold, this solution is $\nn=3$.

Let us summarize the results so far. The configuration of branes in
figure \ref{fig:branes} engineers the theories (\ref{eq:S}) with the quiver in figure \ref{fig:quiver}.
A chain of duality from that brane configuration gives us a space that,
with the help of \cite[Theorem 9.1]{bielawski-dancer},
we recognize as (\ref{eq:ads}), with $B_7$ such that ${\cal M}={\rm Cone}(B_7)$ is exactly the same
space that we had obtained as a moduli space of the same theories ${\cal M}$.

In other words, we have derived a gauge/gravity duality between the theories (\ref{eq:S}) and
the solutions (\ref{eq:ads}), with ${\cal M}={\rm Cone}(B_7)$ defined in (\ref{eq:M}).

\medskip
A similar construction also engineers the theories in section
\ref{sec:orientifold}. Together with the D3 branes stretched along
the $x^6$ direction as in figure \ref{fig:branes}, one now
introduces an O3 projection inverting the direction $3,4,5,7,8,9$
(namely, $\vec x$ and $\vec y$ in (\ref{eq:Mth})). There actually
exist many types of O3 projections (usually called O3$^\pm$,
$\tilde {\rm O}3^\pm$), and we have to specify the one we mean. In
fact, the type of orientifold plane changes as one crosses a brane
\cite{evans-johnson-shapere,hanany-kol}. For example, let us
consider the case in which all the $p_i$ in figure
\ref{fig:branes} are even. Then it is consistent to have the O3 to
be of type O3$^-$ between the $i$--th and $(i+1)$--th fivebrane, and
of type O3$^+$ between the $(i-1)$--th and $i$--th fivebrane. This
configuration engineers the theories considered in section
\ref{sec:orientifold}, with superpotential (\ref{eq:WO}). One can
then follow a similar duality chain as the one we considered in
the case without orientifold. One ends up with a space of the type
we just considered (namely, (\ref{eq:ads}), with ${\cal M}$
defined in (\ref{eq:M})), but further quotiented by the action
$\vec{x}\to - \vec{x}$, $\vec{y}\to - \vec{y}$.

\medskip

Let us now go back to the case without orientifolds.
Just like in \cite{aharony-bergman-jafferis-maldacena}, we can take a limit in which
the gauge group ranks $N_i=N$ are large, and the 't Hooft couplings $\lambda_i= N/k_i$ are large;
in that limit, the appropriate description is in type IIA. One way
to take the $k_i$ large is to rescale them all simultaneously:
\begin{equation}\label{eq:K}
    k_i = K \tilde k_i
\end{equation}
with the $\tilde k_i$ coprime. We already remarked in section \ref{sec:ms} that the
group ${\bf N}$ (defined
in (\ref{eq:N})) has then a discrete component. In fact, it turns out that
\begin{equation}\label{eq:zk}
    {\cal M}(K)= ({\cal M}(K=1))/\zz_K
\end{equation}
where $\zz_K$ acts on $\phi^1$ (recall that the metric on ${\cal M}$ is given
in (\ref{eq:hypertnut}).

We can now take $K\to \infty$. Since the size of $\phi^1$ shrinks to smaller and smaller
size, the solution is best described in IIA:
\begin{equation}\label{eq:IIA}
    {\rm AdS}_4 \times M_6\ , \qquad M_6= B_7/{\rm U}(1) \ .
\end{equation}

Let us give more details about $M_6$.
One  might worry that the U$(1)$ quotient in (\ref{eq:IIA})
might introduce singularities in the IIA solution.

In fact, this does not happen for the particular
U$(1)$ we are quotienting. Let us see how
an orbifold singularity could arise. If,
at a certain locus, the U$(1)$ orbit happened to close for
$\phi^1=\frac{2\pi}k$, for $k$ some integer, then there would be
a $\zz_k$ singularity in the quotient. For a standard example of such
a phenomenon, consider
the sphere $S^{2n-1}$ described by the locus $\sum_{i=1}^n a_i |z_i|^2=1$ in
$\cc^n$, and quotient it by the action $z_i \to e^{i a_i \theta} z_i$. At
the point $z_2=\ldots= z_n=0$, for example, the U$(1)$ action closes
at $\theta= \frac{2 \pi}{a_1}$. The quotient is indeed the weighted projective
space ${\mathbb WP}(a_1, \ldots, a_n)$, which has an orbifold singularity at
the point $z_2=\ldots= z_n=0$.

Now, for us, the loci to worry about are the ones defined in (\ref{eq:deg}).
At each of those, the cycle $(1,p_i)$ of the $T^2$ degenerates. The U$(1)$
action we are considering is along the cycle $(0,1)$. Since $(0,1)$ and
$(1,p_i)$ make up together a $\zz$--basis of $\zz^2$, the U$(1)$ action
does not close before $2\pi$, and there is no orbifold singularity.
Notice that this would not have been the case if we had considered the
diagonal action, along cycle $(1,1)$. This action is in many ways more
natural from a mathematical point of view, and in particular it is part
of the triple of ``Reeb vectors'' that can be used to define a 3--Sasaki structure
\cite[Prop.~1.2.2]{boyer-galicki}. Indeed, it is almost never the case
that one of those vectors yields a quotient without orbifold singularities
(see \cite[Theorem 4.4.5, Prop.~1.2.10]{boyer-galicki}). Fortunately,
we managed to avoid that possible problem.

Physically, we actually remarked earlier that the branes only coincide
at the origin; this is another way of seeing that there should be no
singularity in $M_6$.

Even though we have not studied in detail the topology of $M_6$,
looking at the brane configuration one would expect $n-1$ 2--cycles.
The period of $F_2$ on the $i$--th 2--cycle should then be $p_i$.

We can say more about $M_6$ again in the $n=3$ case. The relevant
${\cal M}$ and $B_7$ were described
in section \ref{sec:example}; when reducing to IIA, we obtain
\begin{equation}
    M_{6\,,n=3}= ({\rm U}(1)\times {\rm U}(1))\backslash {\rm U}(3)/{\rm U}(1)\ .
\end{equation}
Its isometry group is ${\rm SU}(2)\times {\rm U}(1)$, and the cohomogeneity
is 2. Together with the fact that we are not reducing on a Reeb vector
of the 3--Sasakian structure,
this indicates that there is no reason in this case for the dilaton to be
constant. 

\medskip

In conclusion, we have seen that the theories defined in (\ref{eq:S})
are dual to a IIA configuration of the form (\ref{eq:IIA}), with ${\cal M}={\rm Cone}(B_7)$
defined in (\ref{eq:M}).

\bigskip

{\bf Acknowledgments.} We would like to thank O.~Aharony, O.~Bergman,
S.~Giombi, L.~Rastelli, D.~Sorokin, G.~Villadoro, X.~Yin, A.~Zaffaroni
for interesting
discussions. D.~J.~is supported in part by DOE grant DE-FG02-96ER40949, and A.~T.~by DOE grant
DE-FG02-91ER4064.


\providecommand{\href}[2]{#2}

\end{document}